\documentclass[11pt]{article}
\usepackage[dvips]{graphicx}
\newcommand{\be}{\begin{eqnarray}}
\newcommand{\ee}{\end{eqnarray}}

\newcommand{\nd}{\noindent}

\begin{document}
\begin{center}
\medskip
{\large{\bf $J/\psi$ suppression in Au$+$Au collisions at RHIC : colour screening
scenario in the bag model at variable participant numbers}}

\vskip 0.2in

{\large{M. Mishra\footnote{Email: madhukar.12@gmail.com}$^*$, C. P. Singh\footnote{Email: cpsingh\textunderscore bhu@yahoo.co.in}$^\#$, V. J. Menon$^*$ and Ritesh Kumar Dubey$^*$}}\\
{\it {$^*$Dept. of Physics, Banaras Hindu University, Varanasi 221005, India\\
$^\#$Abdus Salam International Centre for Theoretical Physics, 34014 Trieste, Italy}}
\end{center}

\vskip 0.1in

\begin{center}
{\bf Abstract}
\end{center}

\vskip 0.1in

  We have modified the colour screening theory of Chu and Matsui by properly incorporating bag model equation of state for quark gluon plasma (QGP). We have also chosen the pressure parametrization rather than parametrizing energy density in the transverse plane. We assume that the QGP dense medium is expanding in the longitudinal direction obeying Bjorken boost invariant scaling law. Sequential melting of $\chi_c$, $\psi^{'}$ and
$J/\psi$ is also considered in this scenario. We have applied above formulation to the recent PHENIX experimental data of $J/\psi$ suppression in Au$+$Au collisions at RHIC. We find that the model gives a good description of data at mid-rapidity in terms of survival probability versus number of participants without any necessity of implementing (3+1)-dimensional expansion of the deconfined medium.

\vskip 0.9in


\newpage

\section{Introduction}
   One of the main important aims of the Heavy-ion collision
experiments is to detect and find the properties of the state of matter in which quarks and gluons are deconfined as predicted by Lattice Quantum chromodynamics (QCD) above a critical temperature of the order
of $T_c\sim 0.17$ GeV for a baryonic chemical potential $\mu_B=0$ \cite{karsc}. The suppression of heavy quarkonia ($J/\psi,\,\chi_c,\,\psi^{'},\,\Upsilon$) due to colour
screening analogous to Debye charge screening in QED plasma, has long been proposed as a probe of deconfinement in the dense partonic medium. Heavy quarkonia are thus considered as one of the most promising candidates to study the formation and the properties of QGP. In the deconfined state, the interaction between heavy quarks and antiquarks gets reduced due
to colour Debye screening effects leading to a suppression in $J/\psi$ yields. Matsui and Satz~\cite{mats} predict that the binding energy of the $c\bar c$ pair into $J/\psi$ mesons will be screened in
the presence of a QGP medium, leading to the so called $J/\psi$ suppression. In the relativistic heavy ion collisions the $J/\psi$ suppression has been recognized as an important tool to identify possible phase transition from Hadronic matter to quark-gluon plasma (QGP). The $J/\psi$ resonance states are produced at the initial (prethermal) stage of heavy ion collisions because of their large masses. Their small
widths also make them insensitive to final state interactions. Therefore, their evolution probes the deconfined state of matter in the early stage of collisions~\cite{tann}.

   Recently, high statistics data of Au$+$Au collisions at center-of-mass energy $\sqrt{s_{NN}}=200$ GeV at RHIC in Brookhaven National Laboratory have become available~\cite{pher}. It is observed that $J/\psi$ yield in central Au$+$Au collisions is suppressed by a factor of nearly 4 at mid-rapidity and 5 at forward rapidity relative to that in p$+$p collisions scaled by the average number of binary
collisions. Cold nuclear matter (CNM) effects such as nuclear absorption, shadowing and anti-shadowing are also expected to modify the $J/\psi$ yield. CNM effects due to the gluon shadowing and nuclear absorption of $J/\psi$ at the RHIC energy were evaluated from the $J/\psi$ measurement in d$+$Au collisions at RHIC~\cite{ads}. PHENIX d$+$Au data show that CNM effects are smaller at RHIC than at SPS and can be reproduced by considering nuclear absorption cross-section of $J/\psi$ with nucleons and Au nucleus is of the order of $3~mb$ and also incorporating a nuclear-shadowing effect which considers the depletion of low momentum partons in a nucleon embedded in a nucleus compared to their population in a free nucleon. Several groups have measured the $J/\psi$ yield in heavy-ion collisions (for a review of the data and interpretations see the refs.~\cite{vogt,gers}). A $J/\psi$ suppression obtained by the NA50 Collaboration
at SPS~\cite{abreu,nac,aless} could be reproduced by various theoretical models. A larger suppression is expected to occur at RHIC compared to SPS due to a larger energy density present in the medium~\cite{grand,capel}. However, the level of suppression is not much different from that observed by NA50 experiments. Model calculations
assuming colour screening of $J/\psi$ state in a medium depict a much larger suppression at RHIC energies due to the presence of a large parton density, and a higher temperature as well as a large lifetime of the system. On the other hand, several models predict that $J/\psi$ yield will result
from a balance between dissociation~\cite{xusat,gluon1} due to thermal gluons along with colour screening~\cite{chu} and enhancement due to coalescence of uncorrelated $c\bar c$ pair~\cite{grand,pbrau,thews} which are produced abundantly in the initial stage of
collisions at the RHIC energy~\cite{adler,adare}.

   We show in the present work that centrality (i.e., impact parameter or number of participant nucleons $N_{part}$) dependence of the $J/\psi$ suppression in Au$+$Au collisions data at mid-rapidity recently observed by PHENIX experiment at RHIC can be explained invoking a QGP scenario only based on the bag model equation of state (EOS). We demonstrate that the survival probability pattern is well reproduced within present $J/\psi$+hydro model by including the sequential melting of $\chi_c$, $\psi^{'}$ and $J/\psi$ in the longitudinally expanding plasma.
   
%**********************************************************************

\section{Formulation}
Following closely the basic theme of refs.~\cite{chu,dpal} but also highlighting the important differences at appropriate places, our formulation proceeds through the following five stages:

{\nd {\bf{\subsection{Description of the medium}}}}

 In a plasma composed of u, d quarks and gluons, the temperature $T(x)$, energy density $\epsilon(x)$ and pressure $P(x)$ depend on the time-space point $x=(t,\vec{x})$ in the fireball frame. Assuming massless partons and local thermodynamic equilibrium, the bag model equation of state~\cite{john} reads
\begin{eqnarray}
  \epsilon = a\,T^4/c_s^2+B\quad;\quad P=a\,T^4-B\sim a\,(T^4-T_c^4)\\\nonumber
  c_s^2\equiv \frac{\partial P}{\partial\epsilon}\quad;\quad a\equiv \frac{37 \pi^2}{90}
  \quad;\quad B\equiv \frac{17\pi^2\,T_c^4}{45}\sim a\,T_c^4
\end{eqnarray}
where $T_c\sim 0.17$ GeV is the critical temperature for hadron-QGP phase transition,
$c_s^2\sim1/3$ is the square of the velocity of sound in the medium, the coefficient
$a$ contains information about the degrees of freedom, and B is the bag constant. It should be emphasized that above choice of B gives a value $B=0.405$ GeV/fm$^3$. At $T=T_c$ pressure of the QGP medium is zero and this means that QGP behaves as an ideal gas above this temperature.
 It would also be worthwhile to point out that Chu and Matsui~\cite{chu} used bag model equation of state in their colour screening scenario to estimate the proper screening energy density $\epsilon_s$ yet they droped $B$ while defining $c_s^2$ as the ratio $P/\epsilon$. For consistency we shall retain $B$ throughout our analysis in this paper.

The recent lattice QCD simulations do not support well the first order phase transition between the QGP and the hadron gas as is the case we get in the bag model EOS. Still, we want to emphasize that the bag model properly parametrizes many features of the EOS with a rapid change in the entropy density as a function of temperature and hence often continues to be an useful tool employed in the hydrodynamical calculations. If a QGP was produced in central collisions of identical nuclei and is expanding longitudinally, then local thermodynamic observables become function of the lateral coordinate $r$ together with the proper time variable $\tau$ defined by $\tau=(t^2-z^2)^{1/2}$. For algebraic convenience we define the dimensionless ratios
\begin{equation}
  \tilde{\tau}\equiv\frac{\tau}{\tau_i}\quad;\quad \tilde{T}\equiv
  \frac{T(\tau,r)}{T(\tau_i,r)}\quad;\quad \tilde{\epsilon}\equiv\frac{\epsilon(\tau,r)-B}
  {\epsilon(\tau_i,r)-B}\quad;\quad \tilde{P}\equiv\frac{P(\tau,r)+B}{P(\tau_i,r)+B}
\end{equation}
where $\tau_i\sim 1$ fm/c is the proper time for initial thermalization of the
fireball. Then Bjorken differential equation
$\partial\epsilon/\partial\tau=-(\epsilon+P)/\tau$ leads to the scaling solutions or
cooling laws
\begin{equation}
  \tilde{\epsilon}=\tilde{P}={\tilde{T}}^4={\tilde{\tau}}^{-q}\quad;\quad
  q\equiv 1+c_s^2
\end{equation}
The cooling laws written by~\cite{chu,dpal} do not exhibit the effect of $B$ at all.

{\nd {\bf {\subsection{Pressure profile}}}}

Examination of (1) reveals that the pressure almost vanishes at the transition point
$T_c$, i.e., becomes very small in the hadronic sector. This is also supported by the phase diagram given by Blaizot~\cite{blaiz}. Hence on the transverse plane $z=0$ of the fireball and with $t_i=\tau_i$ we
choose the pressure profile
\begin{equation}
  P(t_i,r)=P(t_i,0)\,h(r)\quad;\quad h(r)\equiv \left(1-\frac{r^2}{R_T^2}\right)^{\beta}
  \theta(R_T-r)
\end{equation}
where the coefficient $P(t_i,0)$ is yet to be specified, $R_T$ denotes the radius of
the cylinder and it is related with the transverse overlap area $A_T$ of the colliding
nuclei by $R_T=\sqrt{A_T/\pi}$. The power $\beta\sim 1$ depends on the
energy-deposition mechanism, and $\theta$ is the unit step function, Clearly, our
pressure is maximum at the center of the plasma but vanishes at the edge $R_T$ where
hadronization occurs. In contrast,~\cite{chu,dpal} used a similar parametrization for
the energy density, ignoring the fact that $\epsilon$ should suffer a jump by
$q\,B/c_s^2\sim 4\,B$ across the phase transition point. The factor $P(t_i,0)$ is
related to the mean pressure $<P>_i$ over the cross-section and to the corresponding
average energy density $<\epsilon>_i$ via
\begin{equation}
  P(t_i,0)=(1+\beta)<P>_i=(1+\beta)\{c_s^2<\epsilon>_i-q\,B\}
\end{equation}

Both experimentally and theoretically the determination of $J/\psi$ survival probability S is of paramount importance. This quantity S, in principle, is a function of transverse momentum $p_T$, rapidity $y$, and the impact parameter $b$ (or number of participants $N_{part}$.) Assuming $y\approx 0$ and $b\approx 0$, the original Chu and Matsui model~\cite{chu} was proposed to explain the $p_T$ dependence of the $J/\psi$ suppression in terms of the time dilation of the formation time. However, in the present work our aim is to adopt the dilated formation time concept for describing centrality dependence of the $J/\psi$ suppression data at mid-rapidity which recently became available from PHENIX experiment at RHIC. In order to illustrate this, we take the initial average energy density $<\epsilon>_i$ in terms of the number of participating nucleons $N_{part}$~\cite{khar} (which depends on the impact parameter $b$), given by the modified Bjorken formula:
\begin{equation}
  <\epsilon>_i=\frac{\xi}{A_T\,\tau_i}\frac{dE_T}{dy}\quad;\quad A_T=\pi\,R_T^2
\end{equation}
where $A_T$ is the transverse overlap area of the colliding nuclei and $dE_T/dy$ is
the transverse energy deposited per unit rapidity of output hadrons. Both depend on
the number of participants $N_{part}$~\cite{sscd} and thus provide centrality
dependent initial average energy density $<\epsilon>_i$ in the transverse plane. Next,
$\xi$ is a phenomenological scaling factor discussed latter in Sec. 3 in conjunction
with the self-screened parton cascade model. Of course, \cite{chu,dpal} do not write
any $\xi$ factor in (6) so that their initial average energy density remains
substantially underestimated in numerical applications so that they are also compelled
to employ values determined by the self-screened parton cascade model.

{\nd {\bf{\subsection{Constant pressure contour}}}}

It is well known that a $c\bar{c}$ bound state kept in a thermal medium feels a colour
screened Yukawa potential and it melts at the Debye temperature $T_D$ which corresponds
to the energy density $\epsilon_s$ and pressure $P_s$ given by
\begin{equation}
  T_D\geq T_c\quad;\quad\epsilon_s=a\,T_D^4/c_s^2+B \quad;\quad P_s=a\,T_D^4-B
\end{equation}
For any chosen instant $t$ and on the $z=0$ plane the contour of constant pressure
$P_s$ is obtained by combining the cooling laws (3) with the profile shape (4) to
yield
\begin{equation}
 \tilde{P}\equiv\frac{P_s+B}{P(t_i,0)\,h(r)+B}={\tilde{t}}^{-q}
\end{equation}
Setting $r=0$ the maximum  allowed tilde time $\tilde{t}_{s0}$ (during which pressure
drops to $P_s$ at the center) can be identified as
\begin{equation}
  {\tilde{t}}_{s0}\equiv\left\{\frac{P(t_i,0)+B}{P_s+B}\right\}^{1/q}
\end{equation}
with $P(t_i,0)$ read-off from (5). Thereby the said locus takes the more convenient
form
\begin{equation}
  \left(1-\frac{r^2}{R_T^2}\right)^{\beta}=H_s(t)\equiv\frac{{\tilde{t}}^q-B/(P_s+B)}
  {{\tilde{t}_{s0}}^q-B/(P_s+B)}
\end{equation}
 Our above result generalizes a similar expression derived by~\cite{chu,dpal} for the
$B=0$ case.

{\nd {\bf{\subsection{$J/\psi$ kinemitics and screening radius}}}}

Consider an interacting $c\bar{c}$ created at $t=0$ at the location
$(r_\psi,\phi_\psi)$ on the $z=0$ plane having mass $m_\psi$, momentum $\vec{p}_\psi$,
energy $p_\psi^0=\sqrt{m_\psi^2+p_\psi^2}$, velocity
$\vec{v}_\psi=\vec{p}_\psi/p_\psi^0$, and dilation factor
$\gamma_\psi=p_\psi^0/m_\psi$. In the fireball frame the pair will convert itself into
the physical $J/\psi$ resonance after the lapse of time $t_F=\gamma_\psi\,\tau_F$
(with $\tau_F$ being the intrinsic formation time) provided the temperature $T<T_D$.
From the locus (10) we deduce the so called screening radius
\begin{equation}
  r_s=R_T\,\{1-H^{1/\beta}(t_F)\}^{1/2}\theta\{1-H_s(t_F)\}
\end{equation}
which marks the boundary of the circular region where the quarkonium formation is
prohibited. Hence, the pair will escape the deadly region and form quarkonium if
\begin{equation}
  \mid\vec{r}_\psi+\vec{v}_\psi\,t_F\mid\geq r_s
\end{equation}

Analysis of the kinematic condition is greatly simplified if the $J/\psi$ is moving with pure transverse momentum $p_T$ in the direction (i.e., mid-rapidity domain). Then, for escape the trigonometric condition to be obeyed becomes
\begin{equation}
  cos(\phi)\geq Y\quad;\quad Y\equiv\frac
  {\left[(r_s^2-r^2)\,m-\tau_F^2\,p_T^2/m\right]}{2\,r\,\tau_F\,p_T}
\end{equation}
where all $J/\psi$ suffixes have been omitted without any loss of generality.

{\nd{\bf{\subsection{Survival probability}}}}

Suppose the radial probability distribution for the production of $c\bar{c}$ pair in hard collisions at $r$ is
\begin{equation}
  f(r)\propto\left(1-\frac{r^2}{R_T^2}\right)^{\alpha}\theta(R_T-r)\quad;\quad\alpha\sim 1
\end{equation}
Then, in the colour screening scenario, the net survival probability for the
quarkonium becomes
\begin{equation}
S(p_T)=\frac{\int_0^{R_T} dr\,r\,f(r)\int_{-\phi_{max}}^{\phi_{max}}d\phi}
{2\pi\int_0^{R_T} dr\,r\,f(r)} =\frac{2(\alpha+1)}{\pi\,R_T^2}\int_0^{R_T} dr\,
r\,\phi_{max}(r)\left\{1-\frac{r^2}{R_T^2}\right\}^{\alpha}
\end{equation}
where the maximum positive angle $\phi_{max}$ allowed by (13) is read-off from
\[ \phi_{max}(r) = \left\{\begin{array}{clcr}
            \pi & \mbox{if $Y \leq -1$} \\
            \pi - cos^{-1}\mid Y\mid  & \mbox{if $0\geq Y \geq -1$} \\
            cos^{-1}\mid Y\mid  & \mbox{if $0\leq Y \leq 1$} \\
            0   & \mbox{if $Y \geq 1$}
\end{array}\right.\]

Although the formulae (12-16) exist in the literature, yet we have found and corrected a serious misprint occurring in~\cite{chu,dpal} concerning $\phi_{max}$ in the range $-1\leq Y\leq 0$. In actual practice, it has been found that only about $60\%$ of the observed $J/\psi$ originate directly in hard collisions while $30\%$ of them come from the decay of $\chi_c$ and $10\%$ from the $\psi^{'}$. Hence, the weighted survival
probability of $J/\psi$ becomes
\begin{equation}
  S(p_T)=0.6\,S_{\psi}+0.3\,S_{\chi_c}+0.1\,S_{\psi^{'}}
\end{equation}

%*********************************************************************
\section{Numerical Work}
  Table 1 collects the values of various parameters used in our theory and the following explanations are relevant. The value $T_c=0.17$ GeV is in accord with lattice QCD results. The choice $c_s^2=1/3$ is most common for free massless partons although for partons which carry thermal mass or interact among themselves $c_s^2$ may be different like $1/5$~\cite{dpal}. The selection $\beta=1$ indicates that the energy
deposited in the collision is proportional to the number of nucleon-nucleon
encounters, i.e., to the nuclear thickness. Also, relevant properties of the various quarkonium species in a thermal medium are displayed in Table 2. It is clear that the dissociation temperature $T_D$ gradually decreases in going from $J/\psi$ to $\chi_c$
to $\psi^{'}$.

Our numerical procedure proceeds through the following steps:\\
 (i) Before finding the centrality (or impact parameter) dependence of
$J/\psi$ suppression it is necessary to know the initial average energy density $<\epsilon>_i$ in terms of the number of participants $N_{part}$. For this purpose, we extract the transverse overlap area $A_T$ and the pseudo-rapidity distribution $dE_T/d\eta$ reported in~ref.~\cite{sscd} at various values of  number of participants $N_{part}$. These $dE_T/d\eta$ numbers are then multiplied by a constant Jacobian 1.25 to yield the rapidity distribution  $dE_T/dy$ occurring in (6).\\
(ii) The original Bjorken formula although provides an estimate of the initial energy density qualitatively but, unfortunately, it under-estimates the initial energy density which can cause the suppression of only $\chi_c$ and $\psi^{'}$ but not of
$J/\psi$. Hence, a scaling-up factor $\xi=5$ has been introduced in (6) in order to obtain the desired $<\epsilon>_i=45$ GeV/fm$^3$~\cite{hiran} for most central collision. The appropriate characterization of kinematic quantities in Au$+$Au collisions is presented in Table 3. The relatively large values of our $<\epsilon>_i$ have the following justification:\\
  These are consistent with the predictions of the self-screened parton cascade model~\cite{eskola}, these agree with the requirements of hydrodynamic simulation~\cite{hiran} which fit the pseudo-rapidity distribution of charged particle multiplicity $dN_{ch}/d\eta$ for various centralities already observed at RHIC, and these can cause melting of all the quarkonium species listed in Table 2.\\
(iii) Next, we calculate the time $\tilde{t}_{s0}$ for the pressure to drop to $P_s$ at the
origin and thereby deduce the screening radius $r_s$ with the help of (9, 10).\\
(iv) Next, the quantity $Y$ is computed from (13) which sets the condition for the quarkonium to escape from the screening region, and the limiting values of the $\phi_{max}(r)$ are constructed using equation written just below (15).\\
(v) Finally, the survival probability $S(p_T)$, at specified $p_T$ but varying $N_{part}$ is evaluated by
Simpson quadrature from (15,16).\\
(vi) In order to compare the above analysis with the actual experiments it is necessary to convert the $J/\psi$ suppression data
available in terms of nuclear modification factor $R_{AA}$~\cite{pher} into the accepted def. of survival probability $S(p_T)$~\cite{gunji,leitch,raph} namely
\begin{equation}
  S(p_T)=\frac{R_{AA}}{R_{AA}^{CNM}}
\end{equation}
where $R_{AA}$ is the standard nuclear modification factor and $R_{AA}^{CNM}$ is a contribution to $R_{AA}$ originating from CNM effects constrained by the data of $d+Au$ collisions.

%%%%%%%%%%%%%%%%%%%%%%%%%%%%%%%% begin fig%%%%%%%%%%%%%%%%%%%%%%%%%%%%%%%%%%%%%%%%

\begin{figure}[tbp]
\begin{center}
{\includegraphics[scale=1.0]{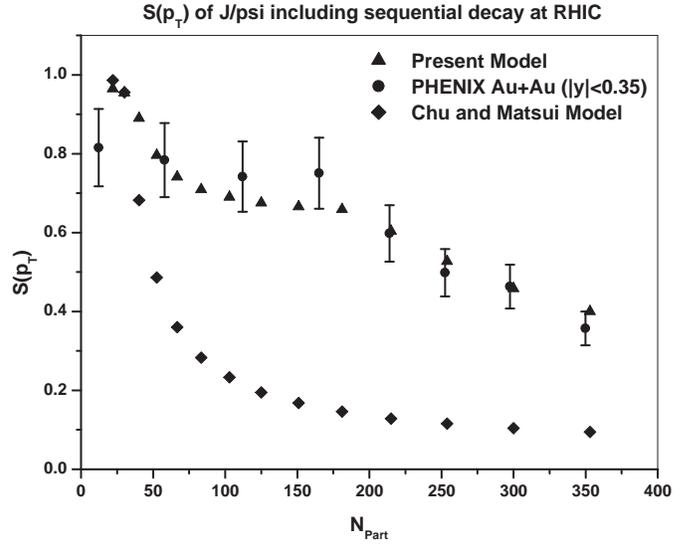}}
\caption{Graph depicting the survival
probability versus number of participants at fixed $p_T=2$ GeV/c allowed by invariant $p_T$ spectrum of $J/\psi$ measured by PHENIX~\cite{pher,gunji,leitch,raph} when sequential melting is considered.}
\end{center}
\end{figure}

\begin{figure}[tbp]
\begin{center}
{\includegraphics[scale=1.0]{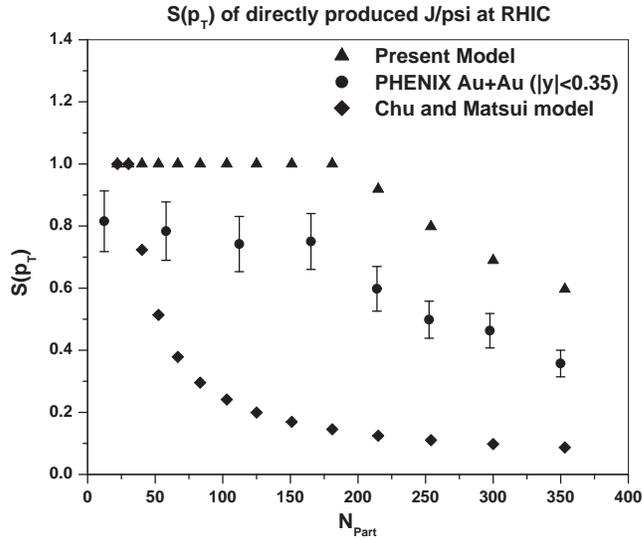}}
\caption{Same as Fig. 1 except that sequential melting is not considered.}
\end{center}
\end{figure}

%%%%%%%%%%%%%%%%%%%%%%%%%%%%%%%% begin tab %%%%%%%%%%%%%%%%%%%%%%%%%%%%%%%%%%%%%%%

\begin{table}[tbp]
\caption{Various parameters used in the theory.}
\begin{center}
\begin{tabular}{llllll}
\hline $T_c$ (GeV) & $c_s^2$ & $B$ (GeV/fm$^3$) & $P_s$ (GeV/fm$^3$) & $\beta$ &
$\alpha$\\ \hline $0.17$ & $1/3$ & $0.405$ & $9.39$ & $1$ & $0.5$\\\hline
\end{tabular}
\end{center}
\end{table}
\begin{table}[tbp]
\caption{Masses, formation times and dissociation temperatures of $J/\psi$, $\chi_c$
nd $\psi^{'}$~\cite{satz}.}
\begin{center}
\begin{tabular}{llll}
\hline \mbox{} & $\psi$ & $\chi_c$ & $\psi^{'}$\\
 \hline $m$ (GeV) & 3.1 & 3.5 & 3.7 \\
$\tau_F$ (fm)& 0.89 & 2.0 & 1.5 \\
 $T_D/T_c$ & 2.1 & 1.16 & 1.12 \\ \hline
\end{tabular}
\end{center}
\end{table}

\begin{table}[tbp]
\caption{Kinematic characterization of Au$+$Au collisions at RHIC~\cite{pher}}
\begin{center}
\begin{tabular}{lll|lll}
\hline Nuclei & $\sqrt{s_{NN}}$ (GeV) & $\xi$ & $N_{part}$ & $<\epsilon>_i$
(GeV/fm$^3$) & $R_T$ (fm)\\ \hline
         &     &     & 22.0 & 5.86 & 3.45 \\
         &     &     & 30.2 & 7.92 & 3.61 \\
         &     &     & 40.2 & 10.14& 3.79 \\
         &     &     & 52.5 & 12.76& 3.96 \\
         &     &     & 66.7 & 15.69& 4.16 \\
         &     &     & 83.3 & 18.58& 4.37 \\
Au$+$Au  & 200 & 5.0 & 103.0& 21.36& 4.61 \\
         &     &     & 125.0& 24.38& 4.85 \\
         &     &     & 151.0& 27.37& 5.12 \\
         &     &     & 181.0& 30.52& 5.38 \\
         &     &     & 215.0& 34.17& 5.64 \\
         &     &     & 254.0& 37.39& 5.97 \\
         &     &     & 300.0& 41.08& 6.31 \\
         &     &     & 353.0& 45.09& 6.68 \\\hline
\end{tabular}
\end{center}
\end{table}
%%%%%%%%%%%%%%%%%%%%%%%%%%%%%%%% end tab             %%%%%%%%%%%%%%%%%%%%%%%%%%%%
Now we turn to physical interpretations of our results.
\section{Results and Discussions}
We present our numerical results under the following two headings:

{\nd \em Sequential decay included.} Figure 1 shows the variation of Survival probability $S(p_T)$ with respect to number of participant nucleons $N_{part}$ at transverse momentum $p_T=2$ GeV/c allowed by the invariant $p_T$ spectrum of $J/\psi$ measured by PHENIX~\cite{pher} at RHIC energy by including sequential decay of $\chi_c$ and $\psi^{'}$ into $J\psi$ calculated from (15, 16). The three curves correspond to experimental data [3; solid circles], our formulation using the bag model
EOS (solid triangles) and old Chu-Matsui theory without bag model EOS [18; solid rhombus]. It is obvious from Figure 1 that $S(p_T)$ decreases with increase in number of participant nucleons in all the three curves because of the growth of the initial
energy density. However, the agreement between experiment and our formulation is very good (characterized by the gradual fall of $S(p_T)$ with $N_{part}$) which justifies our use of the pressure parametrization containing bag constant B. On the contrary, the agreement between experiment and Chu-Matsui theory is poor (as characterized by
sharp fall of $S(p_T)$ with $N_{part}$) which negates their use of the energy parametrization ignoring B.\\
  An important remark must be added at this juncture. Although the quality of fit between our model and experimental data is excellent for central collisions (i.e., relatively large values of the $N_{part}$) it worsens somewhat for non-central collisions (i.e., smaller values of $N_{part}$). This may be due to the fact that our model is based on the assumption of Bjorken boost invariant longitudinal expansion,
which requires cylindrical symmetry about the collision axis. This cylindrical symmetry may be violated up to some extent for most non-central collisions.

{\nd \em Sequential decay excluded.} For the sake of comparison Figure 2 shows similar plots when the theory considers only the suppression of directly produced $J/\psi$'s. The old Chu-Matsui model again shows much larger $J/\psi$ suppression (lower values of $S(p_T)$) for central as well as non-central collisions as compared to experimental results. In contrast, the agreement between the trends (i.e., shapes) of our
theoretical curve and experimental data is much better although their absolute magnitudes differ noticeably for central as well as non-central collisions. Thus, inclusion of the sequential melting of the charmonia~\cite{khar} seems to be essential for fitting the data.

\section{Summary and Conclusions}

We have analyzed centrality dependence of the $J/\psi$ suppression data in Au$+$Au collisions available in terms of survival probability versus number of participants at mid-rapidity from PHENIX experiment at RHIC~\cite{pher,gunji,leitch,raph} by using the modified Chu and Matsui~\cite{chu} model with $J/\psi+$hydro framework. We have simultaneously incorporated several features namely, bag model equation of state, longitudinal expansion of the QGP, pressure parametrization (rather than energy parametrization) in the transverse plane, scaling factor in Bjorken formula to generate higher energy densities (compatible with hydrodynamical simulations), dilated $J/\psi$ formation time and sequential melting of charmonia in the colour screening scenario.

   In conclusion, our present $J/\psi+$hydro formulation presents a very good agreement with the experimental mid-rapidity data of $J/\psi$ suppression versus number of participants observed by PHENIX experiment at RHIC. Of course, a more rigorous treatment of the research problem would require incorporation of additional complications such as  (3+1)-dimensional expansion of the QGP~\cite{dpal,gunji}, EOS built from non-equilibrium fugacities of the partons~\cite{biro}, gluonic dissociation~\cite{xusat,gluon1}, etc. Work on these topics will be taken up in a
future communication.

\section*{Acknowledgements}

M. Mishra and Ritesh Kumar Dubey are grateful to the Council of Scientific and Industrial Research~(CSIR), New Delhi for financial assistance. VJM thanks the University Grant Commission, New Delhi for financial support. C. P. Singh acknowledges financial support and hospitality  of Abdus Salam ICTP, Trieste, Italy where part of the work was done.

\pagebreak

\end{document}